\documentclass[twocolumn,raggedbottom]{revtex4-2}

\usepackage{amsmath,amssymb,bm}
\usepackage{graphicx}
\usepackage{hyperref}
\hypersetup{
colorlinks=false,
pdfborder={0 0 0}
}
\usepackage[utf8]{inputenc}

\DeclareUnicodeCharacter{00AD}{}
\begin{document}

\title{Heliciton-Assisted Chirality-Induced Spin Selectivity from Helical Dirac Current}

\author{Ju Gao}
\email{jugao@illinois.edu}
\author{Fang Shen}%
\affiliation{Department of Electrical and Computer Engineering, University of Illinois at Urbana--Champaign, Urbana, Illinois 61801, USA}

\date{\today}

\begin{abstract}
We develop a quantized screw-mode mechanism for chirality-induced spin
selectivity (CISS). The corresponding quantum, termed a heliciton, is a
screw-symmetric environmental excitation with phase coordinate $\phi-qz$,
longitudinal momentum $\hbar q$, and energy $\hbar\Omega_q$, where $q>0$
and $\Omega_q>0$ denote magnitudes. Its absorption and emission convert the
static local chiral vertex developed in our preceding work into an inelastic
resonant scattering process. In first Born approximation, absorption maps
the $\uparrow,k$ channel to the $\downarrow,k+q$ sideband, whereas emission
maps the $\downarrow,k$ channel to the $\uparrow,k-q$ sideband. The two
outputs inherit the same sampled-current overlap $\mathcal J_\chi(k)$ but
have different ladder factors, final momenta, and resonance detunings.
Writing their spectral factors as $\mathcal S_+(k,q)$ for absorption and
$\mathcal S_-(k,q)$ for emission, thermal averaging gives
\[
P_{\rm sb}(k,q;T)=
\frac{\mathcal S_-(k,q)-\mathcal O(T)\mathcal S_+(k,q)}
{\mathcal S_-(k,q)+\mathcal O(T)\mathcal S_+(k,q)},
\qquad
\mathcal O(T)=
\exp\!\left[-\frac{\hbar\Omega_q}{k_BT}\right].
\]
An isolated emission or absorption resonance yields
$P_{\rm sb}\simeq+1$ or $P_{\rm sb}\simeq-1$, respectively, within the
resolved sideband sector. Reversing the screw handedness interchanges the
spin identities of the two momentum sidebands while leaving their spectral
and occupation weights unchanged, and therefore reverses
$P_{\rm sb}$ at every temperature. At high temperature, absorption and
emission have nearly equal occupation weights; at low temperature,
absorption is exponentially suppressed while emission retains its
spontaneous contribution. Liquid-nitrogen temperature can already produce
a pronounced asymmetry for higher-$\Omega_q$ modes. Thus a real,
spatially resolved Dirac wave carrying a spin-dependent helical conserved
current couples locally to a heliciton and evolves, through deterministic
coupled-wave scattering, into thermally weighted spin- and
momentum-resolved sideband channels, without an ad hoc spin-dependent
potential.
\end{abstract}

\maketitle

\section{Introduction}

Chirality-induced spin selectivity  (CISS) refers to the observation that electron transmission through chiral matter can become strongly spin selective, even in systems where conventional intrinsic spin--orbit coupling appears too small to provide a simple microscopic explanation~\cite{NaamanWaldeck2015,YangEtAl2021,EversEtAl2022}. A microscopic theory must account for two steps. First, it must identify how a chiral structure distinguishes spin-resolved electronic channels at the local interaction level. Second, it must explain how this local distinction becomes a dynamical spin polarization in scattering or transport. Existing approaches have explored effective spin--orbit coupling, orbital polarization, dephasing, vibronic coupling, many-body correlations, and electrode- or interface-induced nonequilibrium effects~\cite{Bloometal2024,Sarkar2025Spinterface,Foo2025MindTheGapCISS}. These studies clarify important aspects of CISS, but a direct route from a local chiral coupling to a resonant spin-polarizing scattering process remains needed.

The local vertex for such a route was identified in our preceding work~\cite{gao2026helicaldiraccurrentlocal}. An exact Dirac electron confined in a cylindrical channel carries a spin-resolved helical conserved-current texture already in the $l=0$ sector. Its charge density carries no orbital winding, but its Dirac current has both longitudinal and azimuthal components, with opposite handedness in the two spin-resolved sectors.

The confined Dirac mode was then coupled to a static screw-symmetric scalar potential,
\begin{equation}
V_\chi(\rho,\phi,z)=V_0 f(\rho)\cos(\phi-qz).
\end{equation}
This perturbation is a scalar chiral potential with a screw phase, not an effective spin--orbit coupling operator. For the transition matrix
\begin{equation}
\left(
\begin{array}{cc}
\langle \psi_{\uparrow k'}|V_\chi|\psi_{\uparrow k}\rangle &
\langle \psi_{\uparrow k'}|V_\chi|\psi_{\downarrow k}\rangle \\
\langle \psi_{\downarrow k'}|V_\chi|\psi_{\uparrow k}\rangle &
\langle \psi_{\downarrow k'}|V_\chi|\psi_{\downarrow k}\rangle
\end{array}
\right),
\end{equation}
the diagonal handedness-preserving kernels vanish, while the off-diagonal handedness-conversion kernels survive. After the longitudinal momentum selection imposed by the screw phase, the selected static kernel is
\begin{equation}
{\bf M}_{\rm sel}(k)=
i\frac{\pi V_0\eta_k}{2ec}J_\chi(k)
\left(
\begin{array}{cc}
0 & 2k-q \\
-(2k+q) & 0
\end{array}
\right),
\label{eq:matrix3}
\end{equation}
where
\begin{equation}
J_\chi(k)=-\int_0^R f(\rho)j_\phi^\uparrow(\rho;k)\rho\,d\rho .
\end{equation}
Here $J_\chi(k)$ is the local geometric overlap between the chiral radial profile and the azimuthal conserved current of the confined Dirac state. It is the kernel through which a scalar chiral perturbation samples the spin-resolved helical-current texture, while the screw phase supplies the longitudinal momentum shift $q$. Thus the coupling converts between opposite helical-current sectors without an external magnetic field or an explicit spin-dependent potential.

The static vertex, however, is not yet a CISS mechanism. A time-independent chiral potential imposes angular and longitudinal selection rules, but it does not exchange energy with the chiral environment, carry an occupation number, acquire a linewidth, or distinguish excitation from de-excitation. It identifies the handedness-conversion kernel, but it does not by itself produce a resonant scattering channel or a spin-polarized outgoing spectrum.

The present paper supplies this missing dynamical step. We promote the screw coordinate that generated the static chiral potential to a quantized helical excitation of the environment. The corresponding quantum, called a heliciton, has wave number $q$, frequency $\Omega_q$, and phase coordinate $\phi-qz$. It can exchange angular structure, longitudinal momentum, and energy with the confined helical Dirac-current electron state. A chiral phonon is one possible material realization of such a mode~\cite{Juraschek2025}; the term heliciton denotes the more general screw-symmetric excitation participating in this local exchange. 

With this quantization, the static handedness-conversion kernel becomes a heliciton-assisted inelastic scattering vertex. Let $n_q$ denote the occupation number of the heliciton mode with wave number $q$. The two elementary conversion channels are
\begin{equation}
|\psi_{\uparrow k};n_q\rangle \longrightarrow |\psi_{\downarrow,k+q};n_q-1\rangle ,
\end{equation}
and
\begin{equation}
|\psi_{\downarrow k};n_q\rangle \longrightarrow |\psi_{\uparrow,k-q};n_q+1\rangle .
\end{equation}
The first channel flips $\uparrow\rightarrow\downarrow$ while annihilating one heliciton quantum; the second flips $\downarrow\rightarrow\uparrow$ while creating one heliciton quantum. Both processes inherit the same sampled-current overlap $J_\chi(k)$ from the static theory, but acquire different final momenta, ladder factors, and resonance denominators.

The first Born outgoing state therefore contains two unconverted incident
channels and two spin-flipped inelastic sidebands. Heliciton absorption
produces a spin-down sideband at $k+q$, whereas heliciton emission produces
a spin-up sideband at $k-q$. The two outputs have different kinematic
weights and resonance detunings. An isolated emission resonance yields
$P_{\rm sb}\simeq+1$, while an isolated absorption resonance yields
$P_{\rm sb}\simeq-1$. Reversing the screw handedness interchanges the angular factors $e^{i\phi}$ and $e^{-i\phi}$ while leaving the longitudinal momentum shifts unchanged. It therefore interchanges the spin identities of the two sidebands and reverses the polarization.

Quantization also distinguishes the thermal availability of the two
processes. Heliciton absorption is weighted by the Bose--Einstein
occupation $n_B(\Omega_q,T)$, whereas emission is weighted by
$n_B(\Omega_q,T)+1$. Their ratio is the Boltzmann factor
$\exp[-\hbar\Omega_q/(k_BT)]$. At high temperature, absorption and emission
have comparable occupation weights. At reduced temperature, absorption is
exponentially suppressed while emission retains its spontaneous
contribution. The effect is strongest for higher-$\Omega_q$ modes and can
already become pronounced near liquid-nitrogen temperature.

The mechanism therefore separates three physical roles. The spatially
resolved Dirac wave supplies a spin-dependent helical conserved-current
texture; the quantized screw-symmetric environment supplies momentum and
energy exchange; and resonance together with bosonic occupation selects and
weights the outgoing sidebands. The preceding work established the static
local chiral vertex. The present work converts that vertex into a resonant,
occupation-controlled CISS mechanism.

\section{Quantized Chiral Mode and Heliciton}

We now promote the static screw coordinate to a quantized environmental mode. The electron remains a confined single-particle Dirac wave, while the chiral structure becomes a dynamical degree of freedom that can exchange momentum and energy with it. The chiral mode is characterized by a screw wave number $q$ and frequency $\Omega_q$. The wave number fixes the screw pitch,
\begin{equation}
q=\frac{2\pi}{Z_\chi},
\end{equation}
where $q>0$ denotes the magnitude of the screw wave number. The phase $\phi-qz$ represents one screw handedness. The opposite enantiomer is obtained by reversing the angular part of the screw phase, $\phi-qz\rightarrow-\phi-qz$, while retaining the same longitudinal wave-number magnitude $q$. For the real static chiral potential, the reversed phase is equivalently represented by $\cos(\phi+qz)$, since
\begin{equation}
\cos(-\phi-qz)=\cos(\phi+qz).
\end{equation}

The frequency may be parametrized by an effective inertia and stiffness,
\begin{equation}\label{eq:Omega_definition}
\Omega_q=\sqrt{\frac{K_q}{M_q}},
\end{equation}
and, for a simple elastic helical mode,
\begin{equation}
\Omega_q=\sqrt{\frac{K_\parallel q^2+K_\perp q^4}{M_q}}.
\end{equation}
Here $M_q$ is the effective inertia of the chiral coordinate, $K_q$ is its effective restoring stiffness, $K_\parallel$ characterizes the longitudinal elastic response, and $K_\perp$ characterizes the chiral restoring rigidity. The detailed dispersion is not needed for the selection rule below; the scattering theory only requires a screw-symmetric mode with wave number $q$ and frequency $\Omega_q$.

The coupled equation is
\begin{equation}
i\hbar\frac{\partial}{\partial t}\Psi({\bf r},t)
=
\left[
\hat H_e+\hat H_{\rm int}(t)
\right]\Psi({\bf r},t),
\end{equation}
where
\begin{equation}
\hat H_e=-i\hbar c\,{\boldsymbol \alpha}\cdot\nabla+\gamma^0mc^2+U(\rho).
\end{equation}
The confining potential is the same as in the static treatment: $U(\rho)=0$ for $0<\rho<R$ and $U(\rho)=U$ for $\rho>R$, with $U>0$.

The time-dependent quantized screw-symmetric interaction is
\begin{equation}\label{eq:quantized_interaction}
\hat H_{\rm int}(t)
=
\frac{V_0 f(\rho)}{2}
\left[
\hat a_q^\dagger e^{i(\Omega_q t+\phi-qz)}
+
\hat a_q e^{-i(\Omega_q t+\phi-qz)}
\right],
\end{equation}
with
\begin{equation}
[\hat a_q,\hat a_q^\dagger]=1,
\qquad
\hat N_q=\hat a_q^\dagger\hat a_q .
\end{equation}
This is the dynamical counterpart of the static screw potential. The factors $e^{\pm i(\phi-qz)}$ retain the same angular and longitudinal selection rules as before, while $e^{\pm i\Omega_q t}$ and the ladder operators allow the chiral environment to lose or gain one quantum of energy $\hbar\Omega_q$. For the opposite enantiomer, the phase
$\Omega_qt+\phi-qz$ is replaced by
$\Omega_qt-\phi-qz$. This reverses the angular factor while retaining the same longitudinal factors and mode frequency.

We call the quantum of this screw-symmetric coordinate a heliciton. A heliciton is a quantized chiral mode whose phase appears as $\Omega_q t+\phi-qz$ in the local Dirac interaction. Through this phase, it carries the angular structure $e^{i\phi}$, energy $\hbar\Omega_q$, and longitudinal momentum $\hbar q$ exchanged with the helical Dirac-current electron state. The same sampled-current overlap $J_\chi(k)$ from the static vertex is retained, but the coupling now changes the heliciton occupation and separates absorption from emission. A molecular torsion, conformational wave, polarization mode, or lattice displacement mode can realize this effective coordinate; in the lattice-displacement case, the mode corresponds to a chiral phonon.

To make the conserved electron--heliciton energy explicit, we transform to the rotating frame of the chiral oscillator and remove the explicit time dependence. In this frame, absorption and emission change the heliciton number, while the total energy of the coupled scattering state is fixed. We write
\begin{equation}
\Psi({\bf r},t)
=
e^{-i{\cal E}t/\hbar}
e^{i(\hat N_q+1/2)\Omega_q t}
\Phi({\bf r}).
\end{equation}
Using
\begin{equation}
e^{-i(\hat N_q+1/2)\Omega_q t}
\hat a_q^\dagger
e^{i(\hat N_q+1/2)\Omega_q t}
=
\hat a_q^\dagger e^{-i\Omega_q t},
\end{equation}
and
\begin{equation}
e^{-i(\hat N_q+1/2)\Omega_q t}
\hat a_q
e^{i(\hat N_q+1/2)\Omega_q t}
=
\hat a_q e^{i\Omega_q t},
\end{equation}
the time-dependent equation then becomes the stationary eigenvalue problem
\begin{equation}
{\cal E}\Phi({\bf r})
=
\left[
\hat H_0+\hat V_\chi
\right]\Phi({\bf r}),
\end{equation}
where the unperturbed electron--heliciton Hamiltonian is
\begin{equation}
\hat H_0
=
-i\hbar c\,{\boldsymbol \alpha}\cdot\nabla
+\gamma^0mc^2
+U(\rho)
+
\left(\hat N_q+\frac{1}{2}\right)\hbar\Omega_q,
\end{equation}
and the time-independent quantized chiral vertex is
\begin{equation}
\hat V_\chi
=
\frac{V_0 f(\rho)}{2}
\left[
\hat a_q^\dagger e^{i(\phi-qz)}
+
\hat a_q e^{-i(\phi-qz)}
\right].
\end{equation}

The additional oscillator term $\left(\hat N_q+\frac{1}{2}\right)\hbar\Omega_q$ is the heliciton energy reservoir. It makes the absorption and emission channels stationary product channels with conserved total electron--heliciton energy ${\cal E}$. The operator $\hat V_\chi$ is the quantized chiral vertex: its screw phases enforce the same handedness-conversion and momentum-selection structure as the static potential, while $\hat a_q$ and $\hat a_q^\dagger$ attach each conversion to a definite heliciton-number change. The annihilation term $\hat a_q e^{-i(\phi-qz)}$ removes one heliciton quantum and translates the electron momentum by $+q$; the creation term $\hat a_q^\dagger e^{i(\phi-qz)}$ adds one heliciton quantum and translates the electron momentum by $-q$. Thus one elementary event generated by $\hat V_\chi$ converts handedness, shifts longitudinal momentum, and exchanges the energy $\hbar\Omega_q$, while the total electron--heliciton energy remains fixed.

The coupling amplitude $V_0$ is now the one-heliciton coupling scale. If the chiral coordinate is denoted by $Q_\chi$, the scalar potential felt by the electron can be expanded to first order as $V_\chi\simeq g_\chi Q_\chi$, where $g_\chi=\partial V_\chi/\partial Q_\chi$ is the electron--heliciton coupling slope. Quantizing the coordinate gives the zero-point amplitude
\begin{equation}
Q_{\rm zpf}
=
\sqrt{\frac{\hbar}{2M_q\Omega_q}},
\end{equation}
so that the single-quantum coupling entering $\hat V_\chi$ is
\begin{equation}
V_0=g_\chi Q_{\rm zpf}.
\end{equation}
Here \(M_q\) is the effective inertia of the chiral coordinate.  For representative molecular-scale values of $M_q$ and $\Omega_q$, $Q_{\rm zpf}$ can lie in the sub-picometer to picometer range.

The natural basis is the product of confined electron Dirac states and heliciton number states,
\begin{equation}
|\psi_{sk};n_q\rangle
=
|\psi_{sk}\rangle\otimes |n_q\rangle,
\qquad
s=\uparrow,\downarrow,
\end{equation}
where
\begin{equation}
\hat N_q|n_q\rangle=n_q|n_q\rangle .
\end{equation}
For the confined $l=0$ electron modes used below, the internal spinor parts are
\begin{equation}
\psi_{\uparrow k}(\rho,\phi,z)
=
N_k
\left(
\begin{array}{c}
J_0(\zeta\rho) \\
0 \\
\eta_k k J_0(\zeta\rho) \\
i\eta_k e^{i\phi}\zeta J_1(\zeta\rho)
\end{array}
\right)
e^{ikz},
\,\,\, 0\leq \rho<R,
\end{equation}
and
\begin{equation}
\psi_{\downarrow k}(\rho,\phi,z)
=
N_k
\left(
\begin{array}{c}
0 \\
J_0(\zeta\rho) \\
i\eta_k e^{-i\phi}\zeta J_1(\zeta\rho) \\
-\eta_k k J_0(\zeta\rho)
\end{array}
\right)
e^{ikz},
\,\,\, 0\leq \rho<R.
\end{equation}
The exterior evanescent continuation and normalization are the same as in the static calculation; the matrix elements below sample only the confined Dirac modes.

In the nonrelativistic regime, the unperturbed product-state energies are
\begin{equation}
{\cal E}_{kn_q}
=
mc^2
+
\frac{\hbar^2(\zeta^2+k^2)}{2m}
+
\left(n_q+\frac{1}{2}\right)\hbar\Omega_q .
\end{equation}
The oscillator term is the energy reservoir that turns the static screw-selected vertex into an inelastic scattering channel. The electron momentum shifts by $\pm q$, the heliciton number changes by one, and the combined electron--heliciton energy remains conserved.

This section establishes the unperturbed product channels $|\psi_{sk};n_q\rangle$ and the quantized interaction $\hat V_\chi$ used for the heliciton-assisted scattering calculation below.

\section{Heliciton-Assisted Born Scattering State and Resonant Sidebands}

The stationary electron--heliciton Hamiltonian defines a multichannel scattering problem. The unperturbed Hamiltonian $\hat H_0$ defines the product channels $|\psi_{sk};n_q\rangle$, and the quantized chiral vertex $\hat V_\chi$ connects them. The outgoing state satisfies
\begin{equation}
|\Psi^{(+)}\rangle
=
|\psi_i\rangle
+
\frac{1}{{\cal E}_{kn_q}-\hat H_0+i\Gamma_q/2}
\hat V_\chi|\Psi^{(+)}\rangle .
\end{equation}
The positive imaginary part imposes the outgoing boundary condition, and $\Gamma_q$ is the linewidth of the heliciton-assisted resonance. In first Born approximation,
\begin{equation}
|\Psi^{(+)}\rangle
\simeq
|\psi_i\rangle
+
\frac{1}{{\cal E}_{kn_q}-\hat H_0+i\Gamma_q/2}
\hat V_\chi|\psi_i\rangle .
\end{equation}
This approximation requires the sideband amplitudes generated by $\hat V_\chi$ to remain perturbatively small.

To display both conversion branches at once, take
\begin{equation}\label{eq:coherent_incident_state}
|\psi_i\rangle
=
\frac{1}{\sqrt{2}}|\psi_{\uparrow k};n_q\rangle
+
\frac{e^{i\theta}}{\sqrt{2}}|\psi_{\downarrow k};n_q\rangle .
\end{equation}
The phase $\theta$ does not affect the sideband probabilities, because the two outgoing sidebands differ in spin, momentum, and heliciton number. The same sideband polarization is obtained from an incoherent equal mixture of incoming $\uparrow$ and $\downarrow$ channels.

The screw phase and ladder operators fix the allowed first-order channels. The annihilation term $\hat a_q e^{-i(\phi-qz)}$ removes one heliciton, shifts the electron momentum by $+q$, and converts $\uparrow$ to $\downarrow$. The creation term $\hat a_q^\dagger e^{i(\phi-qz)}$ creates one heliciton, shifts the electron momentum by $-q$, and converts $\downarrow$ to $\uparrow$. Thus
\begin{eqnarray}\label{eq:scattering_state}
|\Psi^{(+)}\rangle
&=&
c_1|\psi_{\uparrow k};n_q\rangle
+
c_2|\psi_{\downarrow k};n_q\rangle \nonumber \\
&+&
c_3|\psi_{\downarrow,k+q};n_q-1\rangle
+
c_4|\psi_{\uparrow,k-q};n_q+1\rangle.
\end{eqnarray}
The first two terms are the incident spin channels; the last two are the absorption sideband
$|\psi_{\uparrow k};n_q\rangle\rightarrow|\psi_{\downarrow,k+q};n_q-1\rangle$
and the emission sideband
$|\psi_{\downarrow k};n_q\rangle\rightarrow|\psi_{\uparrow,k-q};n_q+1\rangle$.

The incident amplitudes are
\begin{equation}
c_1=\frac{1}{\sqrt{2}},
\qquad
c_2=\frac{e^{i\theta}}{\sqrt{2}} .
\end{equation}
The sideband amplitudes are obtained by projecting the Born correction onto the two final product states:
\begin{equation}
c_3
=
\frac{
\langle \psi_{\downarrow,k+q};n_q-1|\hat V_\chi|\psi_i\rangle
}
{
{\cal E}_{kn_q}-{\cal E}_{k+q,n_q-1}+i\Gamma_q/2
},
\end{equation}
and
\begin{equation}
c_4
=
\frac{
\langle \psi_{\uparrow,k-q};n_q+1|\hat V_\chi|\psi_i\rangle
}
{
{\cal E}_{kn_q}-{\cal E}_{k-q,n_q+1}+i\Gamma_q/2
}.
\end{equation}

Each numerator factorizes into the static Dirac-geometric kernel and a heliciton ladder factor. Let $M_{s's}^{\rm sel}(k)$ denote the channel elements of the selected static kernel in Eq.~\eqref{eq:matrix3}, with $V_0=g_\chi Q_{\rm zpf}$. Then
\begin{equation}
\langle \psi_{\downarrow,k+q};n_q-1|\hat V_\chi|\psi_{\uparrow k};n_q\rangle
=
\sqrt{n_q}\,M_{\downarrow\uparrow}^{\rm sel}(k),
\label{eq:matrix_element_c3_factorized}
\end{equation}
and
\begin{equation}
\langle \psi_{\uparrow,k-q};n_q+1|\hat V_\chi|\psi_{\downarrow k};n_q\rangle
=
\sqrt{n_q+1}\,M_{\uparrow\downarrow}^{\rm sel}(k).
\label{eq:matrix_element_c4_factorized}
\end{equation}
Thus the quantized vertex preserves the local Dirac-current sampling kernel of the static theory, but attaches it to heliciton-number change and an inelastic energy denominator.

Using Eq.~\eqref{eq:matrix3} gives
\begin{equation}
c_3
=-\frac{i}{\sqrt{2}}
\frac{\pi \eta_k}{2ec}
g_\chi Q_q^{(-)}J_\chi(k)
\frac{2k+q}
{-\frac{\hbar^2(2k+q)q}{2m}+\hbar\Omega_q+i\Gamma_q/2},
\end{equation}
and
\begin{equation}
c_4
=\frac{ie^{i\theta}}{\sqrt{2}}
\frac{\pi \eta_k}{2ec}
g_\chi Q_q^{(+)}J_\chi(k)
\frac{2k-q}
{\frac{\hbar^2(2k-q)q}{2m}-\hbar\Omega_q+i\Gamma_q/2}.
\end{equation}

The detunings are defined directly from the corresponding product-state
energy differences. For the spin-down absorption sideband at $k+q$,
\begin{equation}
{\cal E}_{kn_q}-{\cal E}_{k+q,n_q-1}
=
-\frac{\hbar^2(2k+q)q}{2m}
+\hbar\Omega_q
\equiv
-\Delta_+(k,q),
\label{eq:absorption_detuning}
\end{equation}
whereas for the spin-up emission sideband at $k-q$,
\begin{equation}
{\cal E}_{kn_q}-{\cal E}_{k-q,n_q+1}
=
\frac{\hbar^2(2k-q)q}{2m}
-\hbar\Omega_q
\equiv
\Delta_-(k,q).
\label{eq:emission_detuning}
\end{equation}
The amplitude denominators are therefore
$-\Delta_+(k,q)+i\Gamma_q/2$ for absorption and
$\Delta_-(k,q)+i\Gamma_q/2$ for emission. Here
\begin{equation}
Q_q^{(-)}=\sqrt{n_q}\,Q_{\rm zpf},
\qquad
Q_q^{(+)}=\sqrt{n_q+1}\,Q_{\rm zpf}.
\label{eq:Q_ladder_factors}
\end{equation}
In the large-occupation limit, $Q_q^{(+)}\simeq Q_q^{(-)}\equiv Q_q$.

The detunings in Eqs.~\eqref{eq:absorption_detuning} and~\eqref{eq:emission_detuning} show that the heliciton
supplies both the screw momentum and the energy quantum required to turn the
static handedness-conversion kernel into a resonant inelastic scattering
channel.

The result is a four-channel outgoing state with two unconverted incident channels and two spin-flipped inelastic sidebands: a spin-down sideband at $k+q$ and a spin-up sideband at $k-q$. These sidebands have opposite heliciton-number changes and generally different resonance weights. We assume below that $q<k$, so that both sidebands propagate forward. The amplitudes $c_3$ and $c_4$ are then used to calculate the momentum-resolved sideband polarization.

\section{Resonant Sideband Polarization and CISS}

For a one-dimensional outgoing channel with longitudinal momentum $k'$, the sideband spectral weight is proportional to the squared amplitude times the final density of states,
\begin{equation}
W_s(k')\propto |c_s(k')|^2\rho_{\rm 1D}(k').
\end{equation}
In the nonrelativistic regime,
\begin{equation}
E(k')=mc^2+\frac{\hbar^2(\zeta^2+k'^2)}{2m},
\,\,\,
\rho_{\rm 1D}(k')\propto
\left|\frac{dE}{dk'}\right|^{-1}
=
\frac{m}{\hbar^2|k'|}.
\end{equation}
Common constants cancel in the polarization ratio, so
\begin{equation}
W_\downarrow(k+q)\propto \frac{|c_3|^2}{|k+q|},
\qquad
W_\uparrow(k-q)\propto \frac{|c_4|^2}{|k-q|}.
\end{equation}
The sideband-resolved polarization is therefore
\begin{eqnarray}\label{eq:Psb_definition}
P_{\rm sb}
&=&
\frac{W_\uparrow(k-q)-W_\downarrow(k+q)}
{W_\uparrow(k-q)+W_\downarrow(k+q)}
\nonumber \\
&=&
\frac{|c_4|^2/|k-q|-|c_3|^2/|k+q|}
{|c_4|^2/|k-q|+|c_3|^2/|k+q|}.
\end{eqnarray}
This quantity refers to the resonant inelastic sideband sector. If unconverted elastic channels are included in the detected signal, the full transmitted-beam polarization is reduced by that background.

This structure suggests a practical design principle for observing a large CISS polarization: the spin-selective signal should be read out in a momentum- or energy-resolved sideband channel rather than only in the unresolved total transmission. The heliciton-assisted process separates the two spin-flipped
channels into different final momenta, $k-q$ and $k+q$, with different resonance detunings. Selecting one resonant sideband therefore also selects one spin channel, allowing $P_{\rm sb}\simeq \pm 1$. By contrast, an unresolved transmitted beam can be diluted by the elastic channels and by the off-resonant sideband.

Substituting $c_3$ and $c_4$ into Eq.~\eqref{eq:Psb_definition}, the common
factors $g_\chi^2$ and $|J_\chi(k)|^2$ cancel, and the polarization can be
written as
\begin{equation}
P_{\rm sb}(k,q)
=
\frac{
\left[Q_q^{(+)}\right]^2\mathcal S_-(k,q)
-
\left[Q_q^{(-)}\right]^2\mathcal S_+(k,q)
}{
\left[Q_q^{(+)}\right]^2\mathcal S_-(k,q)
+
\left[Q_q^{(-)}\right]^2\mathcal S_+(k,q)
}.
\label{eq:Psb_spectral_factors}
\end{equation}
Here the sideband spectral factors are
\begin{equation}
\mathcal S_-(k,q)
=
\frac{(2k-q)^2}
{|k-q|\left[\Delta_-^2(k,q)+(\Gamma_q/2)^2\right]},
\label{eq:emission_spectral_factor}
\end{equation}
and
\begin{equation}
\mathcal S_+(k,q)
=
\frac{(2k+q)^2}
{|k+q|\left[\Delta_+^2(k,q)+(\Gamma_q/2)^2\right]}.
\label{eq:absorption_spectral_factor}
\end{equation}
The factor $\mathcal S_-(k,q)$ describes the outgoing spin-up
heliciton-emission sideband,
\begin{equation}
|\psi_{\downarrow k};n_q\rangle
\rightarrow
|\psi_{\uparrow,k-q};n_q+1\rangle,
\label{eq:emission_sideband_channel}
\end{equation}
whereas $\mathcal S_+(k,q)$ describes the outgoing spin-down
heliciton-absorption sideband,
\begin{equation}
|\psi_{\uparrow k};n_q\rangle
\rightarrow
|\psi_{\downarrow,k+q};n_q-1\rangle.
\label{eq:absorption_sideband_channel}
\end{equation}
Accordingly, the creation factor $Q_q^{(+)}$ multiplies the emission
spectral factor $\mathcal S_-$, while the annihilation factor $Q_q^{(-)}$
multiplies the absorption spectral factor $\mathcal S_+$.

In the large-occupation limit,
\begin{equation}
Q_q^{(+)}\simeq Q_q^{(-)}\equiv Q_q,
\label{eq:large_occupation_Q}
\end{equation}
Eq.~\eqref{eq:Psb_spectral_factors} reduces to
\begin{equation}
P_{\rm sb}(k,q)
\simeq
\frac{
\mathcal S_-(k,q)-\mathcal S_+(k,q)
}{
\mathcal S_-(k,q)+\mathcal S_+(k,q)
}.
\label{eq:Psb}
\end{equation}

The zeros of the detunings defined in Eqs.~\eqref{eq:absorption_detuning}
and~\eqref{eq:emission_detuning} select the two sideband resonances. If the spin-up emission sideband is
selected,
\begin{equation}
\Delta_-(k,q)=0,
\qquad
\hbar\Omega_q\simeq \frac{\hbar^2(2k-q)q}{2m},
\end{equation}
while the spin-down absorption sideband is off resonance, then
\begin{equation}
P_{\rm sb}\simeq +1.
\end{equation}
If the spin-down absorption sideband is selected,
\begin{equation}
\Delta_+(k,q)=0,
\qquad
\hbar\Omega_q\simeq \frac{\hbar^2(2k+q)q}{2m},
\end{equation}
while the spin-up emission sideband is off resonance, then
\begin{equation}
P_{\rm sb}\simeq -1.
\end{equation}
The limits \(P_{\rm sb}\simeq\pm1\) refer to the polarization ratio within the resolved inelastic sideband sector and do not by themselves imply a large total sideband conversion probability. The absolute sideband yield remains
controlled by the coupling strength, resonance detuning, linewidth, and available density of final states.

Throughout the derivation, $q>0$ and $\Omega_q>0$ denote the magnitudes of
the screw wave number and mode frequency, respectively, and the phase
$\Omega_q t+\phi-qz$ represents one screw handedness. The opposite
enantiomer is represented by the reversed angular screw phase
$\Omega_q t-\phi-qz$. At fixed time, its spatial screw dependence has
the opposite handedness, corresponding to $\cos(\phi+qz)$ up to an
overall temporal phase shift.

This reversal interchanges the angular factors $e^{i\phi}$ and
$e^{-i\phi}$ while leaving the longitudinal factors $e^{-iqz}$ and
$e^{iqz}$ unchanged. Because the two angular factors couple to opposite
spin-resolved helical Dirac-current structures, reversing the angular
selection rule simultaneously reverses the associated spin transition,
without changing the longitudinal momentum shifts or detunings.
Consequently, the $k+q$ absorption sideband and the $k-q$ emission
sideband retain their respective spectral factors
$\mathcal S_+(k,q)$ and $\mathcal S_-(k,q)$, while their spin identities
are interchanged. For otherwise equivalent mode occupation, coupling
strength, and linewidth, the spin-resolved sideband weights are therefore
exchanged, giving
\begin{equation}
\overline P_{\rm sb}(k,q)=-P_{\rm sb}(k,q),
\end{equation}
where the overbar denotes the opposite enantiomer. This enantiosensitive
reversal of the spin-resolved sideband sector is the CISS signature of
the mechanism.

\section{Temperature dependence and occupation-controlled sideband selection}
\label{sec:temperature}

The heliciton-assisted mechanism derived above was formulated for a single quantized screw mode with frequency \(\Omega_q\) and Fock-state occupation number \(n_q\). At finite temperature, the sideband weights are thermally averaged over the occupation of the same bosonic mode. The relevant thermal occupation is the Bose--Einstein factor
\begin{equation}
n_B(\Omega_q,T)
=
\frac{1}{\exp(\hbar\Omega_q/k_BT)-1}.
\label{eq:Bose_occupation}
\end{equation}
This averaging assumes that the mode frequency, linewidth, and electronic
spectral factors are not appreciably modified across the thermally occupied
number states. Accordingly, $\Gamma_q$ and the spectral factors
$\mathcal S_\pm(k,q)$ are treated as temperature independent; any intrinsic
temperature dependence of the heliciton linewidth or frequency would provide
an additional thermal dependence beyond $\mathcal O(T)$. Temperature
therefore changes the sideband weights while leaving the helical-current
coupling vertex unchanged.

The frequency $\Omega_q$ is the same screw-mode frequency introduced in
Eq.~\eqref{eq:Omega_definition} and appearing in the resonance detunings. For a real chiral
molecular structure, the most direct estimate of $\Omega_q$ comes from
measured low-frequency collective modes rather than from a detailed elastic
model. For DNA-like structures, collective angular modes associated with
helical-axis motion occur below about $25~{\rm cm}^{-1}$, while radial
collective modes lie roughly in the range
$50\text{--}110~{\rm cm}^{-1}$~\cite{CoccoMonasson1999}. Using
\begin{equation}
\hbar\Omega_q
=
hc\tilde\nu_q
=
0.12398~{\rm meV}
\left(
\frac{\tilde\nu_q}{1~{\rm cm}^{-1}}
\right),
\label{eq:wavenumber_energy}
\end{equation}
representative heliciton energies are
\begin{equation}
\hbar\Omega_q
\simeq
3.10~{\rm meV},\quad
6.20~{\rm meV},\quad
13.64~{\rm meV},
\end{equation}
corresponding to
\begin{equation}
\frac{\hbar\Omega_q}{k_B}
\simeq
36~{\rm K},\quad
72~{\rm K},\quad
158~{\rm K}.
\end{equation}
These values represent low-, intermediate-, and higher-energy heliciton
branches. Different chiral environments, or different branches of the same
environment, can therefore have substantially different thermal occupations
at the same temperature.

The thermal occupation enters the scattering amplitudes through the heliciton ladder factors,
\begin{equation}
\left[Q_q^{(-)}\right]^2\!=\!
n_B(\Omega_q,T)Q_{\rm zpf}^2,
\,
\left[Q_q^{(+)}\right]^2\!=\!
\left[n_B(\Omega_q,T)+1\right]Q_{\rm zpf}^2.
\label{eq:thermal_ladder_factors}
\end{equation}
Absorption requires an initially occupied heliciton mode and is therefore
proportional to $n_B(\Omega_q,T)$. Emission is proportional to
$n_B(\Omega_q,T)+1$, where the additional unity is the spontaneous
contribution of the quantized screw mode.

Using the sideband spectral factors defined in
Eqs.~\eqref{eq:emission_spectral_factor}
and~\eqref{eq:absorption_spectral_factor}, the thermally averaged sideband
polarization is
\begin{eqnarray}
&&P_{\rm sb}(k,q;T)\nonumber\\
&&=\frac{
\left[n_B(\Omega_q,T)+1\right]\mathcal S_-(k,q)\!-\!
n_B(\Omega_q,T)\mathcal S_+(k,q)
}{
\left[n_B(\Omega_q,T)+1\right]\mathcal S_-(k,q)\!+\!
n_B(\Omega_q,T)\mathcal S_+(k,q)
}.
\label{eq:Psb_temperature}
\end{eqnarray}

For a selected heliciton branch, define the occupation factor
\begin{equation}
\mathcal O(T)
\equiv
\frac{n_B(\Omega_q,T)}
{n_B(\Omega_q,T)+1}
=
\exp\!\left(
-\frac{\hbar\Omega_q}{k_BT}
\right).
\label{eq:occupation_factor}
\end{equation}
The factor $\mathcal O(T)$ is the absorption-to-emission occupation ratio
and, equivalently, the Boltzmann factor relating adjacent heliciton
ladder-state populations. Dividing Eq.~\eqref{eq:Psb_temperature} by
$n_B(\Omega_q,T)+1$ gives
\begin{equation}
P_{\rm sb}(k,q;T)
=
\frac{
\mathcal S_-(k,q)
-
\mathcal O(T)\mathcal S_+(k,q)
}{
\mathcal S_-(k,q)
+
\mathcal O(T)\mathcal S_+(k,q)
}.
\label{eq:Psb_occupation_factor}
\end{equation}

Compared with Eq.~\eqref{eq:Psb}, in which
$\mathcal S_-(k,q)$ and $\mathcal S_+(k,q)$ enter with equal occupation
weighting, Eq.~\eqref{eq:Psb_occupation_factor} introduces
$\mathcal O(T)$ as an asymmetric occupation factor. The spectral factors
$\mathcal S_\pm(k,q)$ contain the geometric, kinematic, and resonant
properties of the two sidebands, while $\mathcal O(T)$ selectively weights
the spin-down absorption branch $\mathcal S_+(k,q)$ relative to the
spin-up emission branch $\mathcal S_-(k,q)$.

The spin labels in Eq.~\eqref{eq:Psb_occupation_factor} refer to the
screw handedness chosen in Eq.~\eqref{eq:quantized_interaction}. For the
opposite enantiomer, reversal of the angular selection rule interchanges
the spin identities of the absorption and emission sidebands while
leaving their spectral factors $\mathcal S_\pm(k,q)$ and occupation
weights unchanged. Therefore, the enantiosensitive reversal holds at
every temperature:
\begin{equation}
\overline P_{\rm sb}(k,q;T)
=
-P_{\rm sb}(k,q;T),
\label{eq:Psb_enantiomer_temperature}
\end{equation}
where the overbar denotes the opposite enantiomer.

In the high-temperature limit,
\begin{equation}
k_BT\gg\hbar\Omega_q,
\qquad
\mathcal O(T)\simeq1,
\label{eq:high_temperature_occupation}
\end{equation}
the two sidebands have equal occupation weighting. The
occupation-induced asymmetry is washed out, and
Eq.~\eqref{eq:Psb_occupation_factor} reduces to the large-occupation result
Eq.~\eqref{eq:Psb}. Any remaining asymmetry is determined by the distinct
kinematic and resonance factors in $\mathcal S_-$ and $\mathcal S_+$.
Both resonance-selected limits therefore remain available,
\begin{equation}
P_{\rm sb}\simeq+1
\qquad\text{or}\qquad
P_{\rm sb}\simeq-1.
\end{equation}

Room temperature,
\begin{equation}
T=300~{\rm K},
\qquad
k_BT\simeq25.9~{\rm meV},
\end{equation}
provides a practical realization of this thermally occupied regime for
sufficiently soft molecular heliciton modes. For the representative
heliciton energies above,
\begin{equation}
\mathcal O(300~{\rm K})
\simeq
0.89,\quad
0.79,\quad
0.59.
\label{eq:occupation_room_temperature}
\end{equation}
The two softer branches therefore have nearly equal absorption and emission
occupation weights, so both sidebands remain readily available. The
higher-energy branch already exhibits an occupation bias, but its
absorption branch is not yet strongly depleted. Room-temperature CISS is
therefore compatible with soft heliciton modes for which $\mathcal O(T)$
remains close to unity, while resonance can select either spin-sideband
output.

As the temperature is lowered, $\mathcal O(T)$ decreases exponentially and
the absorption-selected branch is progressively removed from the sideband
distribution. In the low-temperature limit,
\begin{equation}
\lim_{T\rightarrow0}n_B(\Omega_q,T)=0,
\qquad
\lim_{T\rightarrow0}\mathcal O(T)=0.
\label{eq:low_temperature_occupation}
\end{equation}
For the reference screw handedness used in Eq.~\eqref{eq:quantized_interaction}, Eq.~\eqref{eq:Psb_occupation_factor} gives
\begin{equation}
P_{\rm sb}(k,q;0)=+1,
\label{eq:Psb_zero_temperature}
\end{equation}
provided that the spin-up emission sideband is energetically allowed and
detected. The spin-down absorption contribution
$\mathcal O(T)\mathcal S_+(k,q)$ vanishes, while the spin-up emission
sideband retains the spontaneous contribution. For an exact ground state
$n_q=0$, this follows directly from $\sqrt{n_q}=0$ and
$\sqrt{n_q+1}=1$: absorption is forbidden, whereas emission remains
allowed.

Liquid-nitrogen temperature,
\begin{equation}
T_{\rm LN}\simeq77~{\rm K},
\qquad
k_BT_{\rm LN}\simeq6.64~{\rm meV},
\end{equation}
provides an accessible implementation of the emerging low-temperature
asymmetry. For the representative heliciton energies,
\begin{equation}
\mathcal O(77~{\rm K})
\simeq
0.63,\quad
0.39,\quad
0.13.
\label{eq:occupation_liquid_nitrogen}
\end{equation}
For the highest-energy representative mode, the absorption branch therefore
has only about $13\%$ of the emission occupation weight before the spectral
factors are included. The intermediate-energy branch also exhibits
substantial suppression, whereas the lowest-energy branch remains more
thermally occupied. Liquid-nitrogen temperature can thus produce a
pronounced occupation asymmetry for higher-$\Omega_q$ heliciton branches
without requiring liquid-helium or millikelvin operation.

At high temperature, $\mathcal O(T)\rightarrow1$ and occupation does not
distinguish the two sidebands, allowing both absorption- and
emission-selected channels to contribute at room temperature for soft
heliciton modes. At reduced temperature,
$\mathcal O(T)=\exp[-\hbar\Omega_q/(k_BT)]$ suppresses the spin-down
absorption branch while the spin-up emission branch retains its spontaneous
contribution. The suppression is strongest for higher-$\Omega_q$ modes,
for which liquid-nitrogen temperature can already produce a pronounced
sideband asymmetry.

The occupation asymmetry follows from the bosonic ladder factors and has no
counterpart in a purely static chiral potential. Resonance is not required
for its origin, although it can enhance and spectrally isolate the emission
sideband. Momentum- or energy-resolved measurement of the temperature
dependence of $\mathcal O(T)$ can therefore provide a direct test of the
heliciton mechanism through occupation-controlled spin-sideband selection.

\section{Discussion and Conclusion}

This work completes the static local chiral vertex by promoting the screw
coordinate of the environment to a quantized heliciton. The preceding theory
identified vanishing handedness-preserving kernels and nonvanishing
off-diagonal handedness-conversion kernels. The present theory attaches that
same local overlap to heliciton absorption and emission, thereby adding
conserved energy--momentum exchange, resonance, linewidth, occupation
number, and the four-channel Born scattering structure. The result is a
pair of momentum- and spin-resolved inelastic sidebands whose relative
weights define the observable polarization $P_{\rm sb}$.

The defining CISS signature is the enantiosensitive reversal of this
polarization. Reversing the screw handedness interchanges the spin identities
of the $k+q$ absorption and $k-q$ emission sidebands while leaving
$\mathcal S_\pm(k,q)$ and $\mathcal O(T)$ unchanged, so that $\overline P_{\rm sb}(k,q;T)=-P_{\rm sb}(k,q;T)$ at every
temperature. Bosonic occupation then controls the relative branch
weights: at high temperature, absorption and emission have nearly equal
occupation weighting, whereas at low temperature the absorption branch is
suppressed while spontaneous emission remains. Momentum- or energy-resolved
temperature-dependent sideband spectroscopy therefore provides a direct
test of the mechanism.

The interaction remains scalar. No ad hoc spin-dependent potential,
spin--orbit field, or orbital magnetic-field mediator is introduced. In this
formulation, a real, spatially resolved Dirac wave carrying a spin-dependent
helical conserved current couples locally to a quantized screw-symmetric
environmental mode. The underlying electron--heliciton wave dynamics is deterministic in the
enlarged closed system, while the effective linewidth $\Gamma_q$
phenomenologically represents coupling to additional environmental degrees
of freedom and the observed sideband polarization $P_{\rm sb}$ is thermally
weighted by the statistical occupation of the heliciton mode.
The mechanism
therefore identifies CISS as an enantiomer-reversing, thermally controlled
sideband polarization generated by local geometric coupling between wave
structures.

\bibliography{Helix}

\end{document}